\documentclass[11pt,twoside]{article}
\usepackage{latexsym}
\usepackage{amssymb,amsbsy,amsmath,amsfonts,amssymb,amscd}
\usepackage{mathtools}
\usepackage{graphicx}
\usepackage{pdfsync}
\usepackage{subfigure}
\usepackage{color}
\usepackage{lipsum}
\usepackage[utf8]{inputenc}
\usepackage[T1]{fontenc}
\usepackage{lmodern}

\usepackage{subfig}
\usepackage{caption}
\newcommand{\beq}{\begin{equation}}
\newcommand{\eeq}{\end{equation}}
\newcommand{\bea} {\begin{array}{rl}}
\newcommand{\eea} {\end{array}}
\newcommand{\bepa}{\left\{ \begin{array}{l}}
\newcommand{\eepa} {\end{array}\right.}
\newcommand{\qed}{{ \hfill
                    {\unskip\kern 6pt\penalty 500 \raise -2pt\hbox{\vrule\vbox to 6pt{\hrule width 6pt
                    \vfill\hrule}\vrule} \par}   }}

\usepackage{fancyhdr}

\numberwithin{equation}{section}
\makeatletter
\let\@fnsymbol\@arabic
\makeatother

\makeatletter
\newcommand{\myfootnote}[2]{\begingroup
  \def\@makefnmark{}%
  \addtocounter{footnote}{-1}%
  \footnote{\textbf{#1}: #2}%
  \endgroup}
\makeatother

\title{\Large \bf Effects of space structure and combination therapies on phenotypic heterogeneity and drug resistance in solid tumors}

\author{
Alexander Lorz \thanks{Sorbonne Universit\'es, UPMC Univ Paris 06, UMR 7598, Laboratoire Jacques-Louis Lions, F-75005, Paris, France} \thanks{CNRS, UMR 7598, Laboratoire Jacques-Louis Lions, F-75005, Paris, France} \thanks{INRIA-Paris-Rocquencourt, EPC MAMBA, Domaine de Voluceau, BP105, 78153 Le Chesnay Cedex}
\and Tommaso Lorenzi \footnotemark[1] \footnotemark[2] \footnotemark[3] 
\and Jean Clairambault \footnotemark[3] \footnotemark[1] \footnotemark[2]
\and Alexandre Escargueil \thanks{Sorbonne Universit\'es, UPMC Univ Paris 06, F-75005, Paris, France} \thanks{INSERM, UMR\_S 938, Laboratory of ``Cancer Biology and Therapeutics'', F-75012, Paris, France  \newline Emails: {\fontfamily{pcr} \selectfont alexander.lorz@upmc.fr$,\;$tommaso.lorenzi@upmc.fr$,\;$jean.clairambault@inria.fr, alexandre.escargueil@inserm.fr$,\;\;\;$benoit.perthame@upmc.fr}}
\and Beno\^ \i t Perthame \footnotemark[1] \footnotemark[2] \footnotemark[3] 
}
\date{\today}

\graphicspath{{./Images/}}

\begin{document}

\pagenumbering{arabic}
\maketitle
\pagestyle{plain}

\begin{abstract}
Histopathological evidence supports the idea that the emergence of phenotypic heterogeneity and resistance to cytotoxic drugs can be considered as a process of adaptation, or evolution, in tumor cell populations. In this framework, can we explain intra-tumor heterogeneity in terms of cell adaptation to local conditions? How do anti-cancer therapies affect the outcome of cell competition for nutrients within solid tumors? Can we overcome the emergence of resistance and favor the eradication of cancer cells by using combination therapies? Bearing these questions in mind, we develop a model describing cell dynamics inside a tumor spheroid under the effects of cytotoxic and cytostatic drugs. Cancer cells are assumed to be structured as a population by two real variables standing for space position and the expression level of a cytotoxic resistant phenotype. The model takes explicitly into account the dynamics of resources and anti-cancer drugs as well as their interactions with the cell population under treatment. We analyze the effects of space structure and combination therapies on phenotypic heterogeneity and chemotherapeutic resistance. Furthermore, we study the efficacy of combined therapy protocols based on constant infusion and/or bang-bang delivery of cytotoxic and cytostatic drugs.
\end{abstract}

\section{Introduction}
Cytotoxic drugs are the most widely used weapon in the fight against cancer. However, these drugs usually cause unwanted toxic side effects in the patients' organisms, since they are seldom specific toward tumor cells. Furthermore, they tend to kill strongly proliferative clones, usually considered as made of the most drug-sensitive cells \cite{Mitchison2012}, thus favoring the emergence of resistance to therapies \cite{GoldieColdman1998, Gottesman2002, KomarovaWodarz2005, Szakacs2006}. These are the two major obstacles - toxic side effects and emergence of resistant clones - encountered in the clinic when making use of cytotoxic agents in treating tumors. This situation calls for therapy optimization, that is, identification of drug doses and design of optimal delivery schedules in multi-drug combinations, allowing for an effective control of cancer growth, consisting of a reduction in the probability of resistance emergence together with the minimization of side effects on healthy tissues.

As regards multi-drug combinations, a trend in the modern clinic of cancers leads to combining
cytotoxic drugs, (i.e., DNA damaging agents, antimetabolites, etc.) that lead hit cells to their death, together with cytostatic ones, defined as drugs that are not intended, at least at non massive doses, to harm cells, but rather to slow down proliferation (by blocking growth factor receptors or downstream intracellular pathways involved in proliferation, e.g., tyrosine kinase inhibitors). In fact, cytostatic drugs have lower toxicity for
healthy cells and allow the survival of a small number of cancer clones, that are assumed to be sensitive to cytotoxic agents \cite{SilvaGatenby2010,TomasettiLevy2010,TomasettiLevy22010}. Since sensitive cells can hamper the growth of the resistant ones through competition
for space and resources, this mode of therapy combination allows to attain the twofold goal of
reducing toxicity and holding in check the multiplication of resistant clones, thus establishing as a practical therapeutic strategy the principle at the basis of adaptive therapies: maintaining the persistence of sensitive tumor cells,
which are more fit than the resistant ones in low drug pressure conditions, instead of pursuing the often elusive goal of eradicating the tumor as a whole \cite{Gatenby_2009,GatenbySilvaGillies_2009}.

As far as drug delivery schedules are concerned, it has been suggested that infusion protocols based on bang-bang control (i.e., those protocols in which drug delivery is alternatively
switched on and off over time) can allow an effective control of tumor size \cite{LedzewiczSchattler2002}. We will focus here on the case where tumor cells are exposed to square-wave infusions of cytotoxic and/or cytostatic drugs at constant concentrations and with different durations/maximal doses.

Histopathological evidence supports the idea that the emergence of resistance to anti-cancer
therapies can be considered as a process of Darwinian micro-evolution in tumor cell populations \cite{GerlingerRowanHorswell_etal2012, MerloPepperReid2006}.
In fact, malignant clones with heterogeneous genetic/epigenetic expression leading to different phenotypes (e.g., epithelial vs. mesenchymal, with the same genetic material \cite{Weinberg2007}) can be seen as competing for space and resources (i.e., oxygen, glucose or other nutrients) within the environment defined by the surrounding tissues, together with the selective pressure exerted by therapeutic actions.

According to this view, focusing here on a cancer cell population as reference system (i.e., not taking into account unwanted toxicity to healthy cells, which is a theme we had explored in a previous paper \cite{LorzLorenziHochbergClairambaultPerthame2012}), we propose a structured population model describing cellular dynamics under the effects of cytotoxic and/or cytostatic drugs. The model we design includes birth and death processes involving cancer cells. Furthermore, it also takes explicitly into account the dynamics of resources and anti-cancer drugs as well as their interactions with the cell population under treatment.

Tumor cells are assumed to be organized in a radially symmetric spheroid and to be structured as a population by two non-negative real variables $x \in [0; 1]$ and $r \in [0; 1]$ standing, respectively, for the normalized expression level of a cytotoxic resistant phenotype and for the normalized distance from the center of the spheroid. This implies that, unlike in probabilistic or individual based models, we do not consider that a cell is necessarily either totally sensitive or totally resistant to a given drug; we rather introduce a continuous structuring variable describing resistance between 0 (highly sensitive) and 1 (highly resistant). 

It should be noted that, compared with the model proposed in \cite{LorzLorenziHochbergClairambaultPerthame2012}, the present one is able to mimic the simultaneous selection
of several traits (i.e., the rise of phenotypic polymorphism) within the cancer cell population, which provides the basis for intra-tumor heterogeneity. The additional spatial structure variable $r$, together
with the diffusion along the $r$-axis of nutrients and therapeutic drugs, are the key ingredients of this model that make possible the emergence of such a heterogeneous scenario, which is close to the ones observed in biological experimentations \cite{BuschXingYu_etal2009, GerlingerRowanHorswell_etal2012,Swanton2012}. An alternative way to obtain the emergence of intra-tumor heterogeneity has been namely proposed in \cite{lavi2013role}, by considering sufficiently large mutations in the models from  \cite{LorzLorenziHochbergClairambaultPerthame2012}.   

At this stage, let us stress that both structure variables, $x$ and $r$, have a well defined biological meaning, so that they can be evaluated by means of laboratory experiments. In particular, a cell resistance level can be measured either by the average molecular cell concentration, or, better, activity, of ABC transporters, that are known to be associated with resistance to the drug \cite{Scotto2003, Szakacs2006}, or by the minimal dose of each drug under consideration to kill a given percentage of the cell population \cite{Zhou1996}. 

Let us furthermore mention that the derivation of models able to include both evolution and spatial dynamics, as the one here presented, is a key step toward a better comprehension of those mechanisms that underly the evolution of ecological systems in general. These models can pave the way to interesting mathematical questions; see for instance \cite{MirrahimiRaoul2013} and references therein. 

The paper is organized as follows. In Section 2, we describe the mathematical model, the related underlying assumptions and the general setup for numerical simulations. Section 3 is devoted to study cell environmental adaptation in the framework of this model, i.e., how tumor cells adapt to the surrounding environment defined by nutrients and anti-cancer drugs. In particular, the evolution of phenotypic heterogeneity and chemotherapeutic resistance are analyzed in presence of cytotoxic and cytostatic drugs. With the aim of supporting the design of optimal therapeutic strategies, in Section 4 we test the efficacy of therapeutic protocols based on constant infusion and/or bang-bang delivery (i.e., infusion schedules relying on bang-bang control) of cytotoxic drugs, cytostatic drugs or combinations of cytotoxic and cytostatic drugs. Conclusions are drawn in Section 5, which provides also some ideas about future research perspectives.

\section{A structured population model for a tumor cell spheroid exposed to anti-cancer drugs}
\label{sec:mod2}
We first present the mathematical model and the related underlying assumptions. In more detail, Subsection \ref{sec:mod2A} describes the strategies developed to translate into mathematical terms the phenomena under consideration and presents the model, while Subsection \ref{sec:mod2B} summarizes the general setup for numerical simulations of the Cauchy problem defined by endowing the model with biologically consistent initial and boundary conditions.
\subsection{Model and underlying assumptions}
\label{sec:mod2A}
The reference system is defined by a population of tumor cells exposed to cytotoxic and cytostatic drugs. As previously noted, the cell population is assumed to be organized in a radially symmetric spheroid and to be structured by two non-negative real variables $x \in [0,1]$ and $r \in [0,1]$. The former represents the normalized expression level of a cytotoxic resistant phenotype (i.e., roughly speaking, the level of resistance to cytotoxic agents), while the latter stands for the distance of cells from the center of the spheroid, whose radius is assumed to be normalized in order to have unitary length.

The density of cancer cells is modeled by function $n(t,r,x) \geq 0$, so that local and total density at time $t \in [0,\infty)$ are computed, respectively, as
$$
\varrho(t,r) = \int_0^1 n(t,r,x) dx, \qquad \varrho_T(t) = \int_0^1 \varrho(t,r) dr,
$$
while the average level of resistance $\chi(t)$ expressed by the whole cell population and the related variance $\sigma^2(t)$ can be evaluated as
$$
\chi(t) = \int_0^1 \int_0^1 x \frac{n(t,r,x)}{\varrho_T(t)} dx dr, \qquad \sigma^2(t) = \int_0^1 \int_0^1 x^2 \frac{n(t,r,x)}{\varrho_T(t)} dx dr - \chi(t)^2.
$$
In the mathematical framework at hand, function $\sigma^2(t)$ provides a possible measure for intra-tumor phenotypic heterogeneity at time $t$.
Function $s(t,r) \geq 0$ identifies the concentration of nutrients available to cells (oxygen and glucose, since in this setting we do not consider the glycolytic phenotype, i.e., we do not distinguish between these two nutrients). The densities of cytotoxic and cytostatic drugs are described, respectively, by $c_{1}(t,r) \geq 0$ and $c_2(t,r) \geq 0$.

We assume that the evolution of functions $n$, $s$, $c_{1}$ and $c_2$ is ruled by the following set of equations
\begin{equation}\label{MOD2}
\partial_t n(t,r,x) = \left[ \frac{p(x)}{1+\mu_2 c_2(t,r)} s(t,r) - d(x)\varrho(t,r) - \mu_1(x)c_1(t,r) \right] n(t,r,x),
\end{equation}
\begin{equation}
- \sigma_s \Delta s(t,r) + \left[ \gamma_s + \int_0^1 p(x) n(t,r,x) dx\right] s(t,r) = 0,
\end{equation}
\begin{equation}
- \sigma_c \Delta c_{1}(t,r) + \left[ \gamma_{c} + \int_0^1 \mu_{1}(x) n(t,r,x) dx\right] c_{1}(t,r) = 0,
\end{equation}
\begin{equation}
- \sigma_c \Delta c_{2}(t,r) + \left[ \gamma_{c} + \mu_{2} \int_0^1 n(t,r,x) dx\right] c_{2}(t,r) = 0,
\end{equation}
with zero Neumann conditions at $r=0$ coming from radial symmetry and Dirichlet boundary conditions at $r=1$
\begin{equation}\label{BC-1}
s(t,r=1) = s_1, \quad \partial_r s(t,r=0)=0, \quad c_{1,2}(t,r=1)=C_{1,2}(t), \quad \partial_r c_{1,2}(t,r=0)=0,
\end{equation}
where:
\\\\
$\bullet$ Function $p(x)$ is the proliferation rate of cells expressing resistance level $x$ due to the consumption of resources. Factor
$$
 \frac{1}{1+\mu_2 c_2(t,r)}
$$
mimics the effects of cytostatic drugs, which act by slowing down cellular proliferation, rather than by killing cells. Parameter $\mu_2$ models the average uptake rate of these drugs.
\\
$\bullet$ Function $d(x)$ models the death rate of cells with resistance level $x$ due to the competition for space and resources with the other cells.
\\
$\bullet$ Function $\mu_1(x)$ denotes the destruction rate of cells due to the consumption of cytotoxic drugs, whose effects are here summed up directly on mortality (i.e., in this simple setting, not involving the cell division cycle, we do not consider drug effects on cell cycle phase transitions \cite{KimmelSwierniak2006}).
\\
$\bullet$ Parameters $\sigma_s$ and $\sigma_c$ model, respectively, the diffusion constants of nutrients and cytotoxic/cytostatic drugs.
\\
$\bullet$ Parameters $\gamma_s$ and $\gamma_{c}$ represent the decay rate of nutrients and cytotoxic/cytostatic drugs, respectively.
\\\\
Model \eqref{MOD2} can be recast in the equivalent form
$$
\partial_t n(t,r,x) = R\big(x,\varrho(t,r),c_{1,2}(t,r),s(t,r)\big)n(t,r,x),
$$
in order to highlight the role played by the net growth rate of cancer cells, which is described by
$$
R\big(x,\varrho(t,r),c_{1,2}(t,r),s(t,r)\big) := \frac{p(x)}{1+\mu_2 c_2(t,r)} s(t,r) - d(x)\varrho(t,r) - \mu_1(x)c_1(t,r).
$$
The following considerations and hypothesis are assumed to hold:
\\\\
$\bullet$ With the aim of translating into mathematical terms the idea that expressing cytotoxic resistant phenotype implies resource reallocation (i.e., redistribution of energetic resources from proliferation-oriented tasks toward development and maintenance of drug resistance mechanisms, such as higher expression or activity of ABC transporters \cite{Scotto2003, Szakacs2006} in individual cells), we assume $p$ to be decreasing
\begin{equation} \label{B3}
p(\cdot) > 0, \qquad p'(\cdot) < 0.
\end{equation}
$\bullet$
In order to include the fact that mutations conferring resistance to cytotoxic therapies may also provide cells with stronger competitive abilities, function $d$ is assumed to be non-increasing 
\begin{equation} \label{B4}
d(\cdot) > 0, \qquad d'(\cdot) \leq 0.
\end{equation}
$\bullet$ The effects of resistance to cytotoxic therapies are modeled by assuming function $\mu_{1}$ to be non-increasing
\begin{equation} \label{B5}
\mu_{1}(\cdot) > 0, \qquad \mu'_{1}(\cdot) \leq 0.
\end{equation}

\subsection{Setup for numerical simulations}
\label{sec:mod2B}
Numerical simulations are performed in {\sc Matlab} making use of an implicit-explicit finite difference scheme combined with a shooting method with $200 \times 200$ points on the square $[0,1] \times [0,1]$. Interval $[0,T]$ is selected as time domain, with $T=700$ in Section 3 and $T=3000$ in Section 4 (time step $dt=0.1$).

We choose the initial and boundary conditions to be
\begin{equation}\label{BC00}
n(t=0,r,x) = n^0(r,x) := C^0 \exp(-(x-0.5)^2/0.005),
\end{equation}
\begin{equation}\label{BC01}
s(t,r=1) = s_1:=0.3,  \quad c_{1,2}(t,r=1)=C_{1,2}(t),
\end{equation}
where $C_{1,2}(t)$ are positive real functions, which model the infusion rates of cytotoxic/cytostatic drugs and are defined case by case according to the situation under investigation in each subsection of Section 3 and Section 4. Choice \eqref{BC00} mimics a biological scenario where most of the cells are characterized by the same intermediate level of resistance to therapies at the beginning of observations (i.e., the cell population is almost monomorphic). The normalization constant $C^0$ is set equal to 0.1.

In accordance with assumptions \eqref{B3}-\eqref{B5}, the other functions and  parameters of the model are set as follows along all simulations:
\begin{equation}\label{S1}
p(x) := 0.1 + (1-x)  , \qquad d(x) := 3+1.5(1-x)^2, \qquad \mu_1(x) := 0.01 + (1-x)^2,
\end{equation}
and
\begin{equation}\label{S2}
\mu_2 := 10, \qquad \sigma_s=\sigma_c := 0.2, \qquad \gamma_s = \gamma_c := 1.
\end{equation}
The above polynomial functions and the related parameters are chosen to be simple and offering clear illustrations of the generic properties set in \eqref{B3}-\eqref{B5}.

\section{Study of cell environmental adaptation and phenotypic heterogeneity}
\label{sec:mod3}
For the model described in the previous section, we now study how tumor cells adapt to the surrounding environment defined by nutrients and anti-cancer drugs. Subsection \ref{sec:mod3A} deals with cell dynamics without therapies, while in Subsection \ref{sec:mod3B} we analyze the effects of constant infusions of cytotoxic and cytostatic drugs. Considerations about the evolution of intra-tumor heterogeneity are drawn in Subsection \ref{sec:mod3Eb}, and a qualitative mathematical justification for phenotypic selection is provided by Subsection \ref{sec:mod3Db}. 

\subsection{Cell dynamics without therapies}
\label{sec:mod3A}
\begin{figure}[h!]
\centerline{\includegraphics[width=0.8\textwidth]{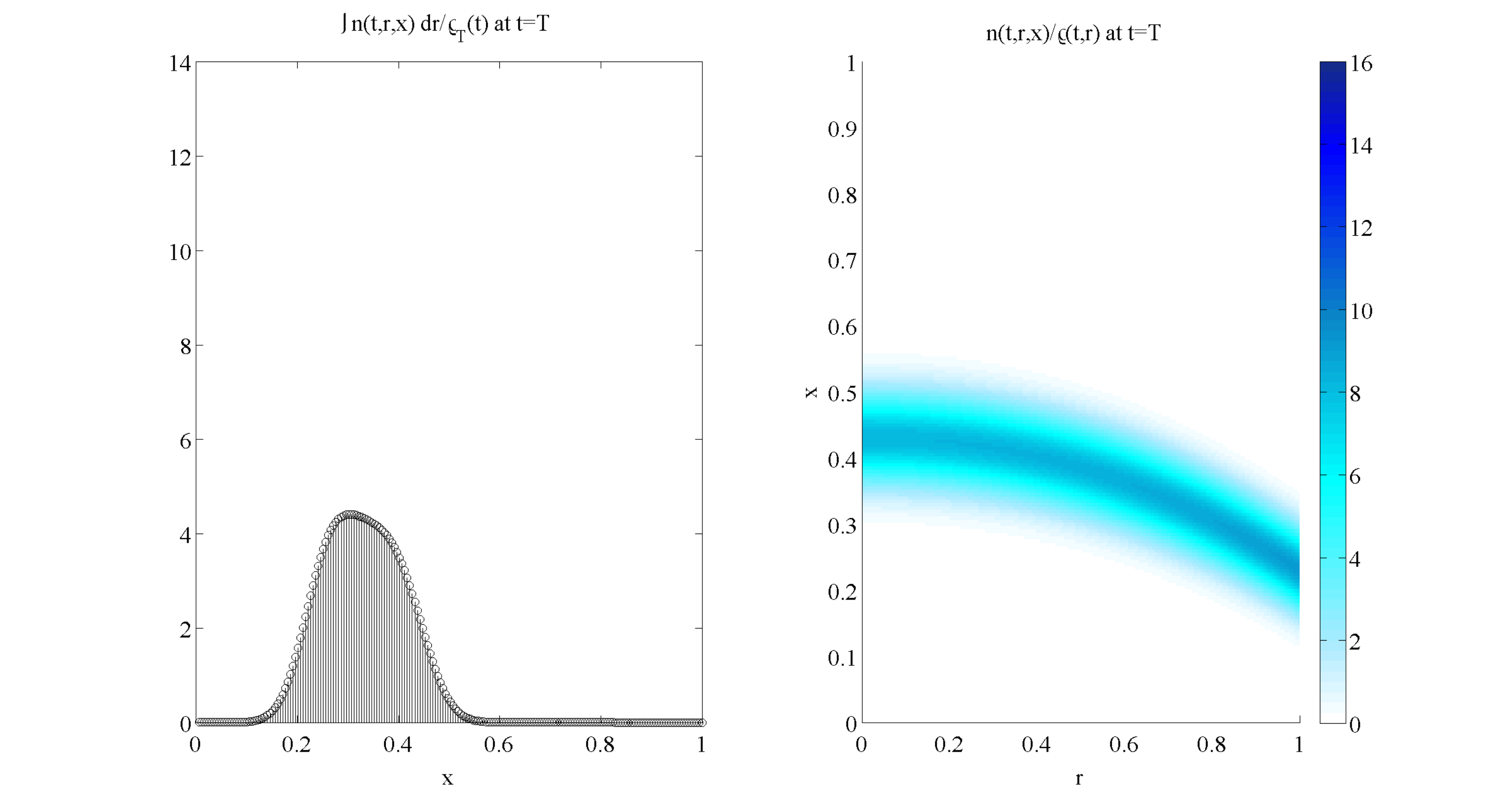}}
\caption{\label{F3} \textbf{(Cell dynamics without therapies)} Plots of the average resistance trait distribution $\displaystyle{\int_0^1 n(T,r,x) dr/\varrho_T(T)}$ (left panel) and the phenotype distribution along the tumor radius  $\displaystyle{n(T,r,x)/\varrho(T,r)}$ (right panel) for $C_{1,2}(t) := 0$. For each $r$ value, the $n(T,r,x)$ function concentrates in a different point $X(T,r)$. Cells characterized by a low expression level of resistance to cytotoxic therapies and by a strong proliferative potential are selected, and this is particularly obvious at the rim of the spheroid ($r=1$), where nutrients abound.}
\end{figure}

\begin{figure}[h!]
\centerline{\includegraphics[width=0.8\textwidth]{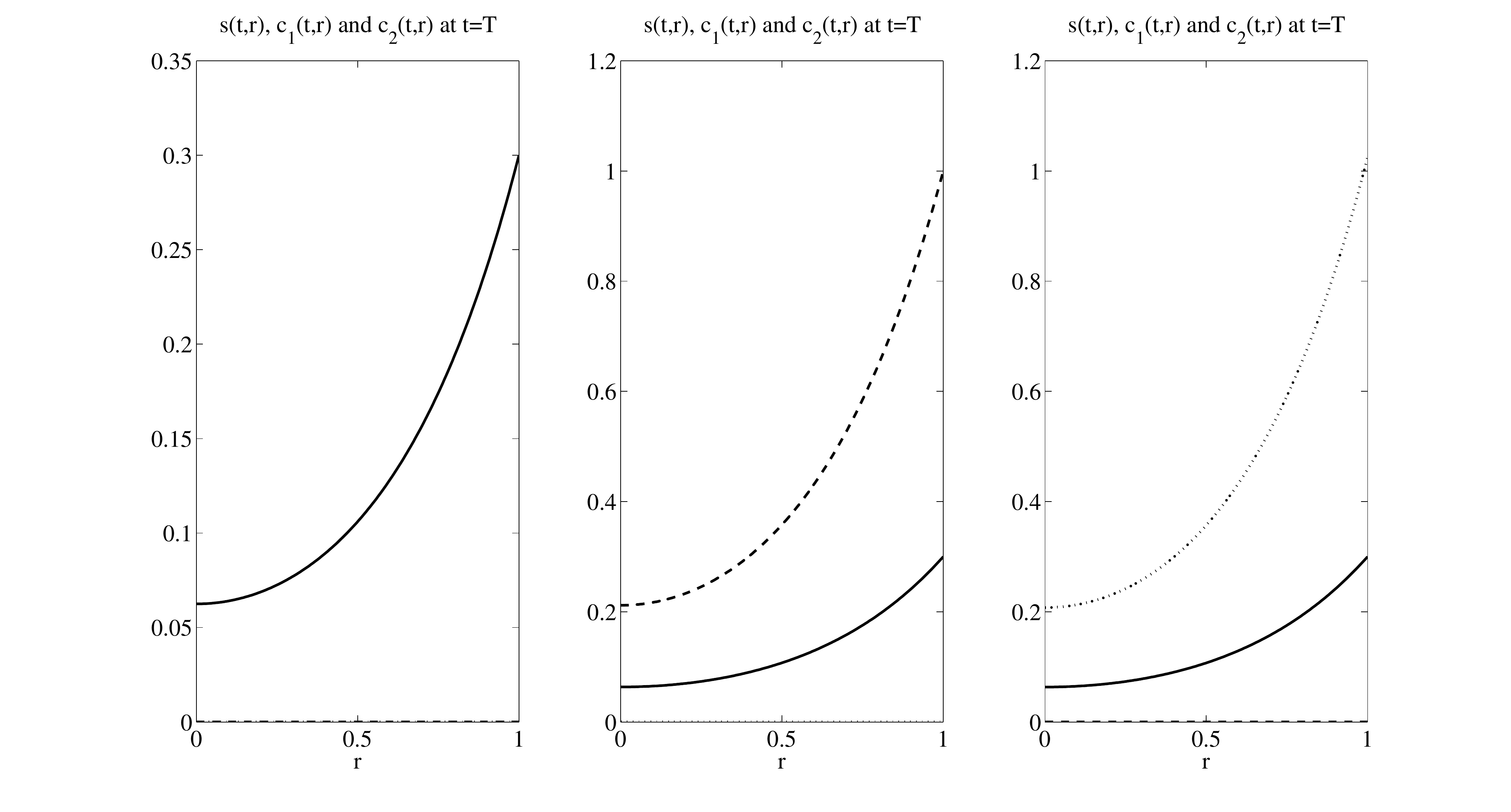}}
\caption{\label{F3A}  \textbf{(Distributions of resources and drugs)} Plot of $s(T,r)$ (solid lines), $c_1(T,r)$ (dashed lines) and $c_2(T,r)$ (dotted lines) for $C_{1,2}(t) := 0$ (left panel), $C_{1}(t) := 1, C_{2}(t) := 0$ (center panel) and $C_{1}(t) := 0, C_{2}(t) := 1$ (right panel).}
\end{figure}

We begin our study by analyzing the dynamics of cancer cells without therapies (i.e., $C_{1,2}(t) := 0$ for any $t \in [0,T]$). The obtained results are summarized by the right panel in Fig.~\ref{F3}, which shows how, in the absence of therapeutic agents, cells characterized by lower resistance levels and thus, using assumption \eqref{B3}, by stronger proliferative potentials, are selected. At each position $r$, a different trait $X(T,r)$ is favored (i.e., for each value of $r$, $n(T,r,x)$ concentrates in a different point $X(T,r)$). This is due to the fact that the density of resources varies along the radius of the spheroid (i.e., $s(T,r)$ attains different values at any $r$, see solid line in the left panel of Fig.~\ref{F3A}). In other words, different densities of available nutrients imply the selection of different levels of abilities to get resources and this provides the basis for the emergence of polymorphism within the tumor cell population at hand.

\subsection{Cell dynamics under infusion of cytotoxic or cytostatic drugs}
\label{sec:mod3B}
\begin{figure}[h!]
\centerline{\includegraphics[width=0.8\textwidth]{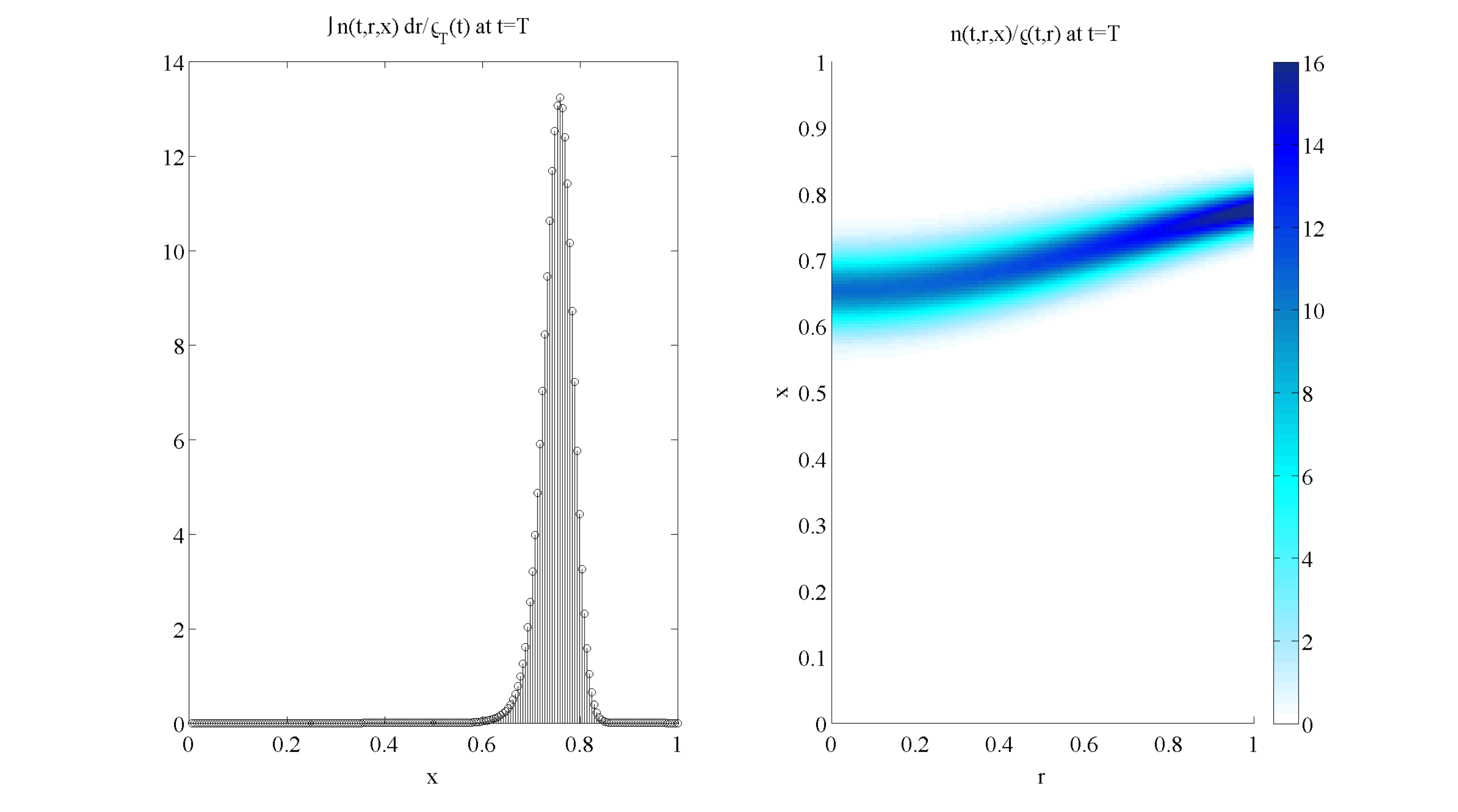}}
\caption{\label{F4} \textbf{(Cell dynamics under constant infusion of cytotoxic drugs)}  Plots of the average resistance trait distribution  $\displaystyle{\int_0^1 n(T,r,x) dr/\varrho_T(T)}$ (left panel) and the phenotype distribution along the tumor radius $\displaystyle{n(T,r,x)/\varrho(T,r)}$ (right panel) for $C_{1}(t) := 1$ and $C_{2}(t) := 0$. For each value of $r$, function $n(T,r,x)$ concentrates in a different point $X(T,r)$. Cells characterized by high resistance levels are selected. As in the case without drugs, such evolution is particularly obvious at the rim of the spheroid ($r=1$), where drugs abound.}
\end{figure}

At first, we consider the effects that constant infusions of cytotoxic drugs induce on cell dynamics, i.e., we run simulations setting $C_1(t) := 1$ and $C_2(t) := 0$ for any $t \in [0,T]$. The right panel in Fig.~\ref{F4} highlights how cytotoxic drugs promote a selective sweep toward resistant phenotypes. A polymorphic scenario arises at the end of simulations also in this case; in fact, a different level of resistance $X(T,r)$ is selected at any level within the spheroid (i.e., for each value of $r$, $n(T,r,x)$ concentrates in a different point $X(T,r)$). By analogy with the case without therapies, this is due to the fact that the concentrations of nutrients and cytotoxic drugs vary along the radius of the spheroid (i.e., $s(T,r)$ and $c_1(T,r)$ attain different values for different values of $r$, see solid and dashed lines in the center panel in Fig.~\ref{F3A}).

In order to study how cancer cells respond to the on-off switch of cytotoxic drug infusion, we develop simulations setting
\begin{equation}\label{C1}
C_1(t):=
\left\{
\begin{array}{lr}
1, \quad \mbox{if } t \in [0,T]
\\
0, \quad \mbox{if } t \in (T,2T],
\end{array}
\right.
\end{equation}
and  keeping $C_2(t):=0$ for any $t \in [0,2T]$. The above definition mimics a biological scenario where cytotoxic drugs are delivered in the time interval $[0,T]$ only. Fig.~\ref{F7} highlights the selection of higher levels of resistance during the infusion of cytotoxic drugs, i.e., in the $[0,T]$ time interval, and higher level of proliferative potential in the absence of cytotoxic drugs, i.e., on the $(T,2T]$ time interval. In fact, when the infusion of cytotoxic drugs is stopped,  more proliferative, and thus less resistant, cancer clones are favored \cite{Gatenby_2009,GatenbySilvaGillies_2009}. The switch from the selection for resistance to proliferative potential occurs in a progressive and continuous way, rather than through jumps in the distribution over the traits (see the right panel in Fig.~\ref{F7}).

\begin{figure}[h!]\label{F7}
\centerline{\includegraphics[width=0.75\textwidth]{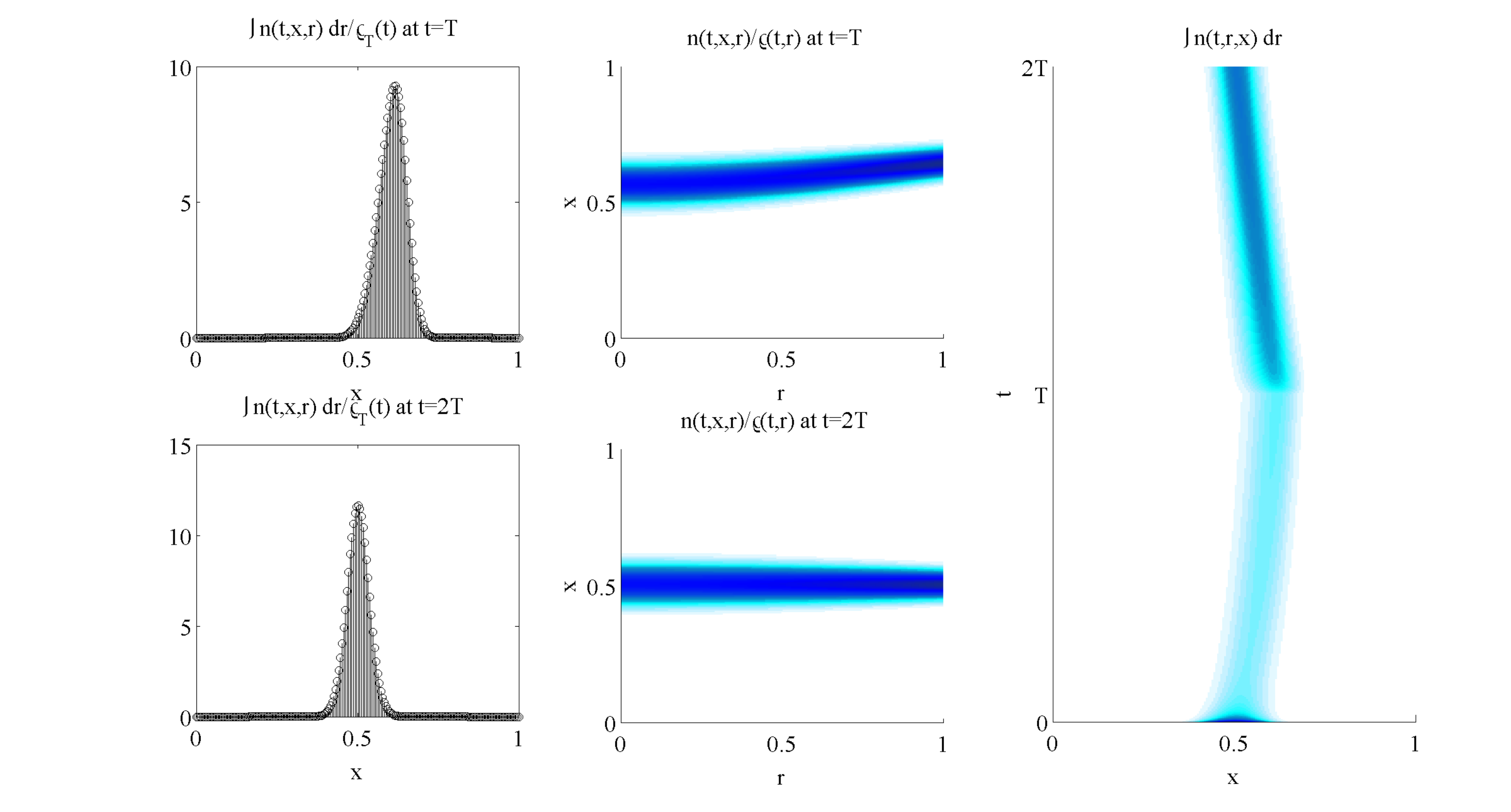}}
\caption{\label{F7} \textbf{(On-off switch of cytotoxic drug infusion)} Plots of $\displaystyle{\int_0^1 n(T,r,x) dr/\varrho_T(T)}$ (left-top panel), $\displaystyle{\int_0^1 n(2T,r,x) dr/\varrho_T(2T)}$ (left-bottom panel), $\displaystyle{n(T,r,x)/\varrho(T,r)}$ (center-top panel), $\displaystyle{n(2T,r,x)/\varrho(2T,r)}$ (center-bottom panel) and $\int_0^1 n(t,r,x) dr$ for $t \in [0,2T]$ (right panel), for $C_1(t)$ defined by \eqref{C1} and $C_2(t):=0$. The selection of higher levels of resistance occurs during the infusion of cytotoxic drugs, i.e., in time interval $[0,T]$, while higher levels of proliferative potential are selected in the absence of cytotoxic drugs, i.e., in time interval $(T,2T]$.}
\end{figure}

\begin{figure}[h!]
\centerline{\includegraphics[width=0.8\textwidth]{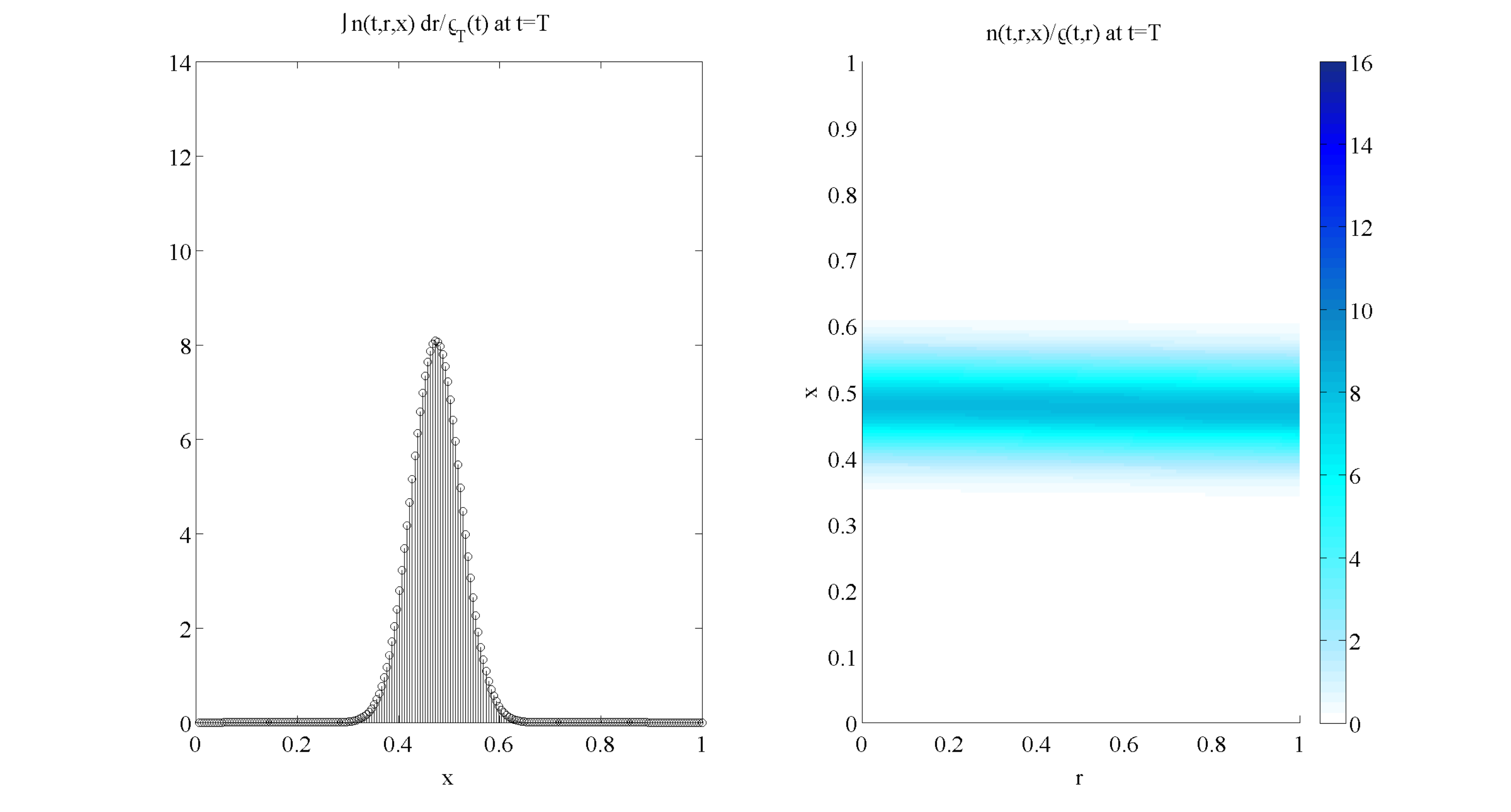}}
\caption{\label{FN1} \textbf{(Cell dynamics under constant infusion of cytostatic  drugs)} Plots of the average resistance trait distribution  $\displaystyle{\int_0^1 n(T,r,x) dr/\varrho_T(T)}$ (left panel) and the phenotype distribution along the tumor radius $\displaystyle{n(T,r,x)/\varrho(T,r)}$ (right panel) for $C_{1}(t) := 0$ and $C_{2}(t) := 1$. Cytostatic drugs slow down the evolution of cancer cells and do not favor the emergence of resistance}
\end{figure}

We subsequently analyze the dynamics of cancer cells under the effects of constant infusion of cytostatic drugs, i.e., we run simulations setting $C_1(t) := 0$ and $C_2(t) := 1$ for any $t \in [0,T]$. The right panel in Fig.~\ref{FN1} shows how the cell distribution at the end of simulations is still close to the initial one, that is, cytostatic drugs tend to slow down the evolution of cancer cells and do not favor the emergence of resistance.
\\\\

{\it The results presented in this subsection lead us to conclude that phenotypic heterogeneity within solid tumor aggregates might come from cell adaptation to local conditions. Cells characterized by different levels of proliferative potential and resistance to therapies are selected depending on space position, in relation with the distributions of resources and anti-cancer drugs. Cytostatic drugs tend to slow down tumor evolution, while cytotoxic drugs favor the selection of highly resistant cancer clones.}

\subsection{Considerations about intra-tumor heterogeneity}
\label{sec:mod3Eb}
\begin{figure}[h!]
\centerline{\includegraphics[width=0.8\textwidth]{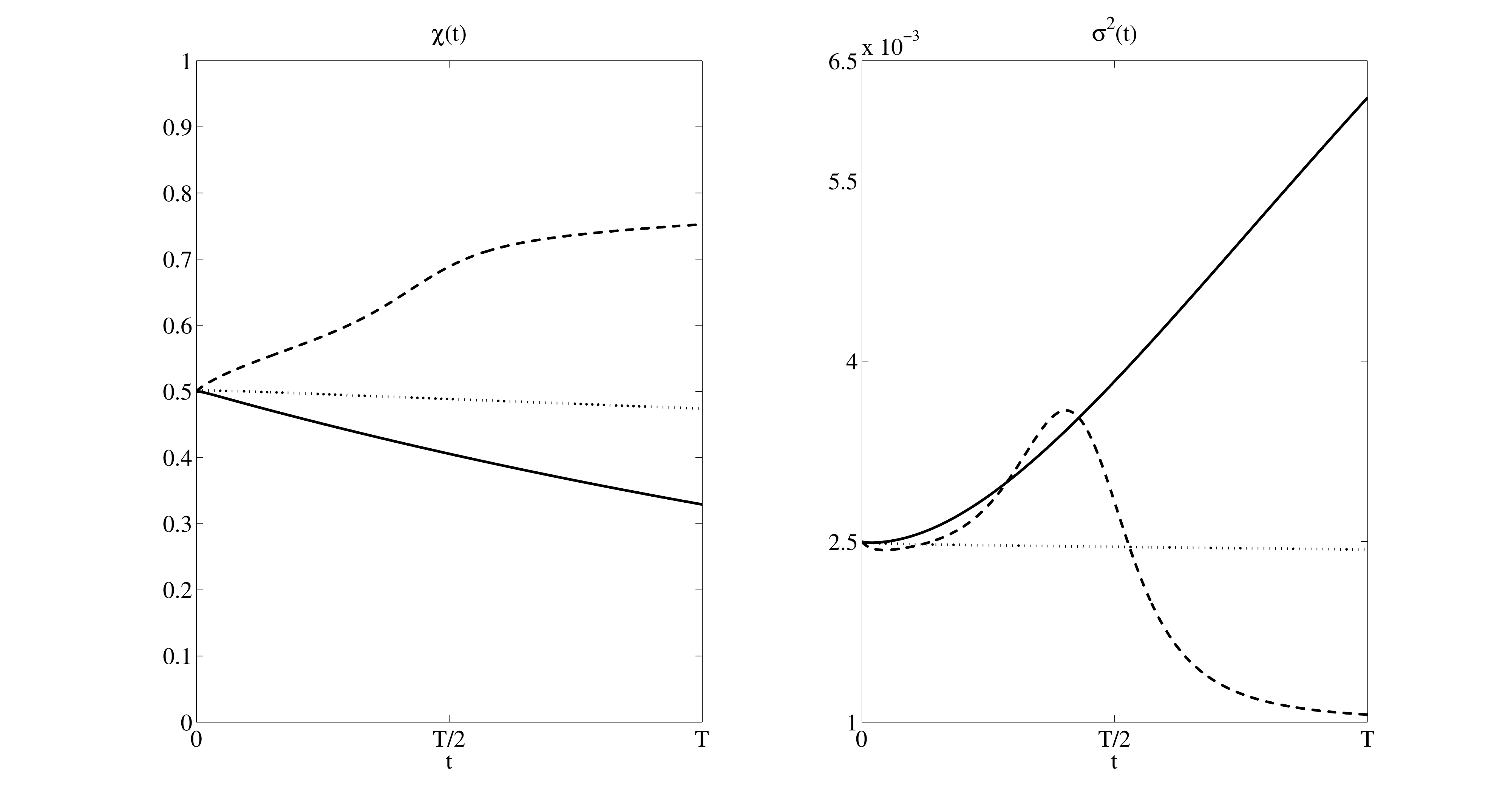}}
\caption{\label{F3h}  \textbf{(Evolution of the average level of resistance and the related variance)} Plot of $\chi(t)$ (left panel) and $\sigma^2(t)$ (right panel) for $C_{1,2}(t) := 0$ (solid lines), $C_{1}(t) := 1, C_{2}(t) := 0$ (dashed lines) and $C_{1}(t) := 0, C_{2}(t) := 1$ (dotted lines). Cytotoxic drugs  increase the average level of resistance $\chi(t)$ over time, while the related variance $\sigma^2(t)$ decreases. This may be interpreted as a reduction of intra-tumor heterogeneity w.r.t. the resistance trait, due to the delivery of the drugs inducing such resistance.}
\end{figure}

A comparison between the results illustrated in the left panels in Fig.~\ref{F3}, Fig.~\ref{F4} and Fig.~\ref{FN1} lead us to conclude that intra-tumor heterogeneity is reduced under the effects of cytotoxic drugs. The same idea is also supported by the results presented in Fig.~\ref{F3h}, which highlight how these drugs increase the average level of resistance expressed by the whole cell population $\chi(t)$ over time, while the related variance $\sigma^2(t)$ decreases. In the framework of our model, this is in agreement with the Gause competitive exclusion principle and it is consistent with experimental observations for the fact that cytotoxic drugs increase the selective pressure and favor highly resistant cancer clones \cite{Gottesman2002, Szakacs2006}.
\\

{\it In conclusion to this study of cell environmental adaptation, we observe that cytotoxic drugs reduce intra-tumor heterogeneity of the resistance trait. This can be seen as an evolutionary bottleneck in the cancer cell population \cite{Gerlinger2010, Swanton2012}}.\\
\newpage
\subsection{A qualitative mathematical justification for phenotypic selection}
\label{sec:mod3Db}
From a mathematical standpoint, taking advantage of the considerations drawn in \cite{LorzMirrahimiPerthame2010,GB.BP:08},  the long term dynamics of $X(t,r)$ can be formally characterized by solving the equation
\begin{equation}\label{RR}
\lim_{t \rightarrow \infty} R\big(x=X(t,r),\varrho(t,r),c_{1,2}(t,r),s(t,r)\big)= R\big(x=\bar{X}(r),\bar{\varrho}(r),\bar{c}_{1,2}(r),\bar{s}(r)\big) = 0.
\end{equation}
In the case at hand, this is equivalent to finding the two roots of a second degree polynomial and verifying whether they belong to the  interval $[0,1]$. Defining
\begin{eqnarray}
&&b(r) = \frac{\bar{s}(r)}{\left[\bar{c}_1(r)+1.5 \bar{\varrho}(r)\right] \left[1+10  \bar{c}_2(r)\right]} - 2,
\nonumber\\
&& c(r) = 1 - \frac{1}{\bar{c}_1(r)+1.5 \bar{\varrho}(r)} \left[\frac{1.1 \bar{s}(r)}{1+10 \bar{c}_2(r)} - 3 \bar{\varrho}(r)-0.01\bar{c}_1(r)\right],
\nonumber
\end{eqnarray}
we verify, through numerical inspection, that the zero of equation \eqref{RR} in the interval $[0,1]$ without drugs and with cytostatic drugs only, is given as
\begin{equation}\label{XAS1}
\bar{X}(r) = \frac{-b(r) + \sqrt{b(r)^2 - 4c(r)}}{2},
\end{equation}
while the zero of equation \eqref{RR} in the interval $[0,1]$  with cytotoxic drugs only, is given as
\begin{equation}\label{XAS2}
\bar{X}(r)  = \frac{-b(r) - \sqrt{b(r)^2 - 4c(r)}}{2}.
\end{equation}
Finally, we need $b(r)^2-4c(r) \ge 0$ otherwise \eqref{RR} does not have a solution.

\begin{figure}[h!]
\centerline{\includegraphics[width=0.8\textwidth]{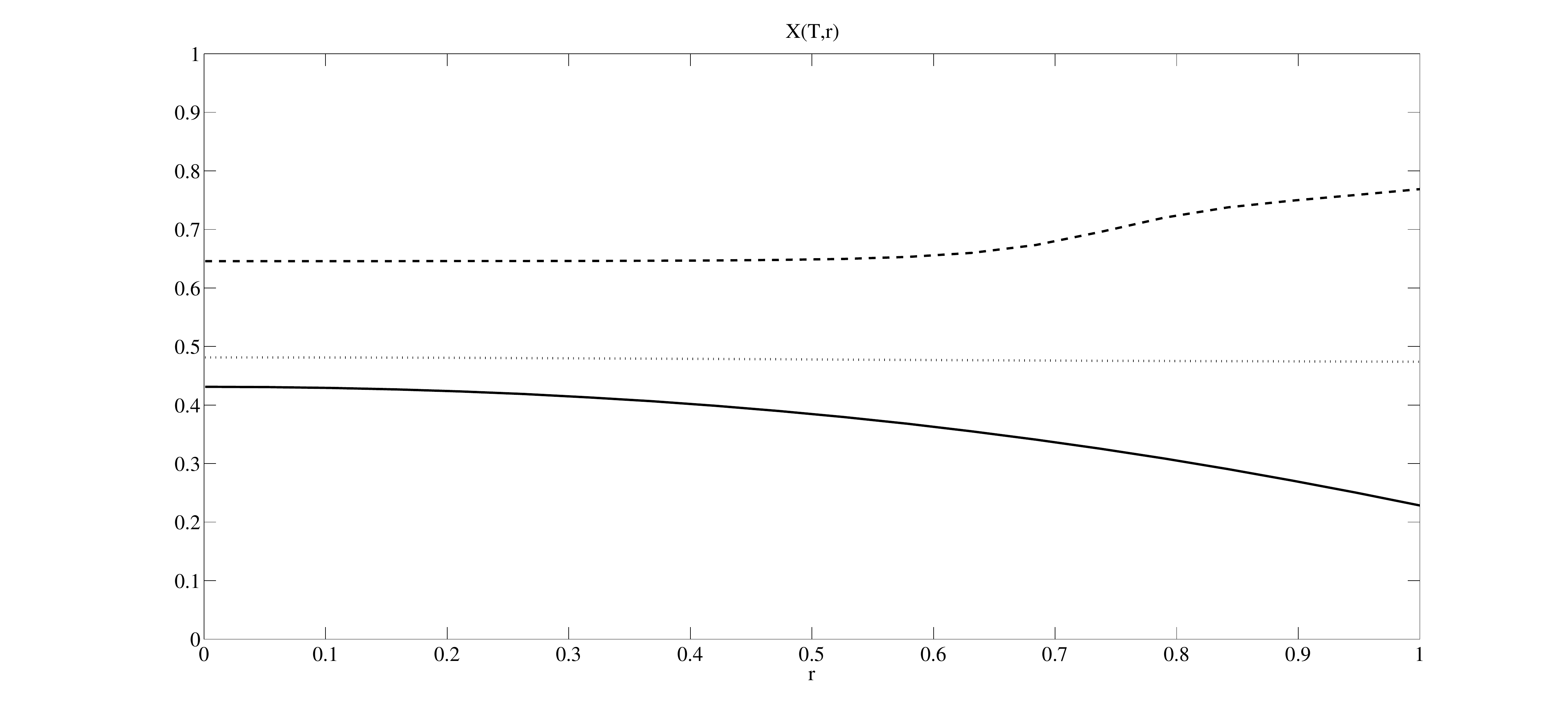}}
\caption[caption]{\label{F5}
\textbf{(Curves $X(T,r)$ computed from equation \eqref{RR})}
Curves $X(T,r)$ for $C_{1,2}(t) := 0$ (solid line), $C_1(t) := 1$ and $C_2(t) := 0$ (dashed line), and $C_1(t) := 0$ and $C_2(t) := 1$ (dotted line).
}
\end{figure}

The curves $X(T,r)$ in Fig.~\ref{F5} obtained from the formula above show a good agreement with the plots of the function $n(T,r,x)/\varrho(T,r)$ summarized in Fig.~\ref{F3}, Fig.~\ref{F4} and Fig.~\ref{FN1}.

\section{Study of optimized therapeutic protocols}
\label{sec:4}
\begin{figure}[h!]
\centerline{\includegraphics[width=0.8\textwidth]{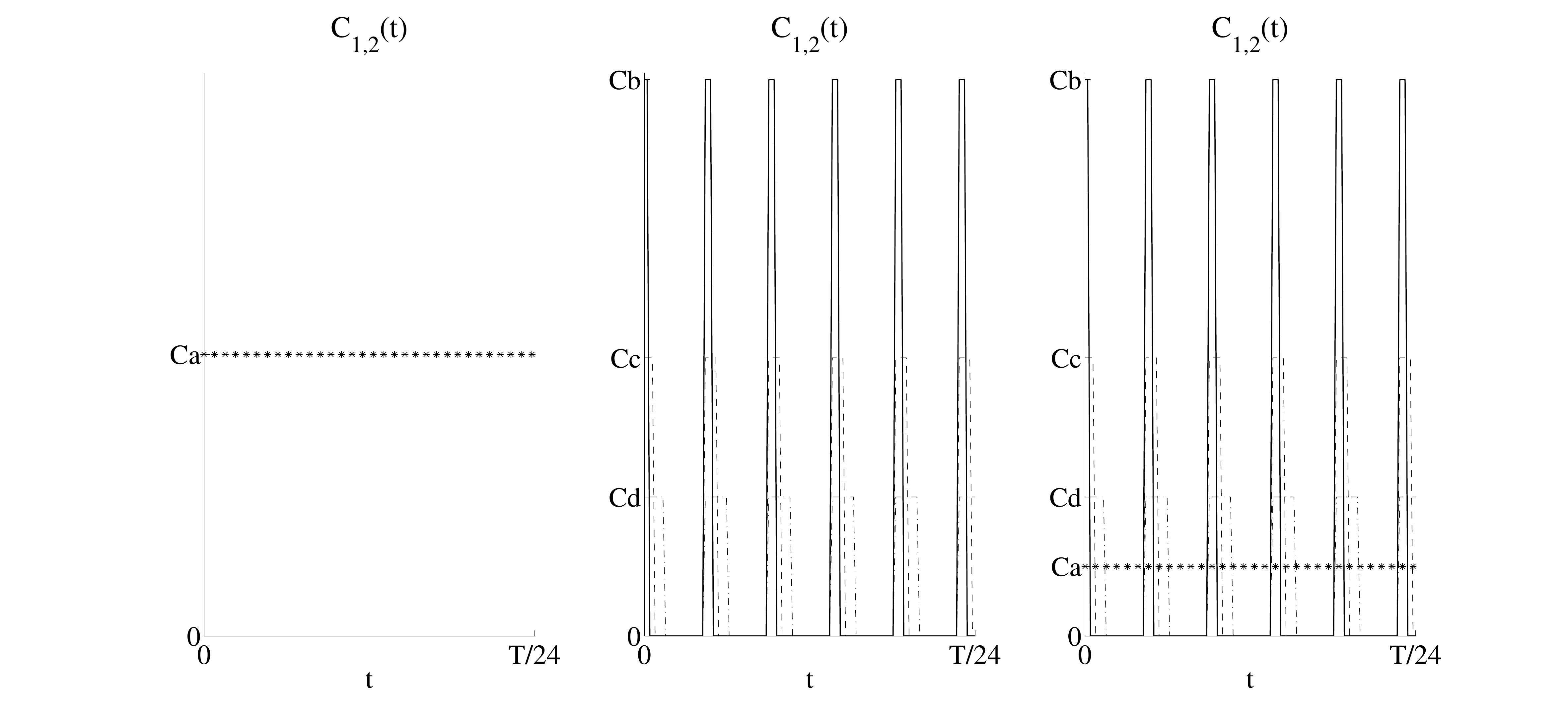}}
\caption[caption]{\label{IS}
\textbf{(Infusion schemes of cytotoxic and cytostatic drugs)} Definitions of boundary conditions $C_{1,2}(t)$. Left and center panels refer to constant and bang-bang infusion of cytotoxic and/or cytostatic drugs, while right panel refers to the case where cytotoxic drugs are delivered through a bang-bang infusion scheme while cytostatic drugs are constantly supplied, or vice-versa. Fixed, dashed and dashed-dotted lines stand for three different instances of bang-bang infusion, which are characterized by picks of different height/duration. The values of constants $C_{a,b,c,d}$ are defined, case by case, according to the situation considered in each subsection.} 
\end{figure}

In this section, we compare the efficacy of different schedules of drug delivery with the aim of supporting the development of optimized therapeutic protocols. The effects of bang-bang infusion of cytotoxic or cytostatic drugs are compared to the ones of constant supply in Subsection \ref{sec:41n}, while the same kind of comparison for cytotoxic and cytostatic drugs in combination is provided in Subsection \ref{sec:41}. Finally, the effects of therapeutic strategies that combine constant delivery of cytotoxic drugs with bang-bang infusion of cytostatic drugs, and vice-versa, are inspected in Subsection \ref{sec:43}. The infusion schemes (i.e., the boundary conditions $C_{1,2}(t)$) in use throughout this section are summarized by Fig.~\ref{IS}, where constants $C_{a,b,c,d}$ model the delivered doses and are defined, case by case, according to the scenario analyzed in each subsection and in such a way that the total delivered dose (i.e., $\int_0^T \left[C_1(t) + C_2(t) \right] dt$) remains always the same. 

\subsection{Constant vs bang-bang infusion of cytotoxic OR cytostatic drugs only}
\label{sec:41n}
At first, we study the efficacy of therapeutic protocols relying on bang-bang delivery of cytotoxic drugs only, and we compare the obtained results with the ones of constant infusion. We perform simulations with $C_2(t):=0$ and $C_1(t)$ defined by the fixed line in the center panel in Fig.~\ref{IS} with $C_b=16$, or the dashed line with $C_c=8$ or the dashed-dotted line with $C_d=4$. The obtained results are compared to the outcomes of simulations developed with $C_2(t):=0$ and $C_1(t)$ defined by the $*$-line in the left panel in Fig.~\ref{IS} with $C_a=2$. 

\begin{figure}[h!]
\centering
\subfigure[]{\includegraphics[width=0.8\textwidth]{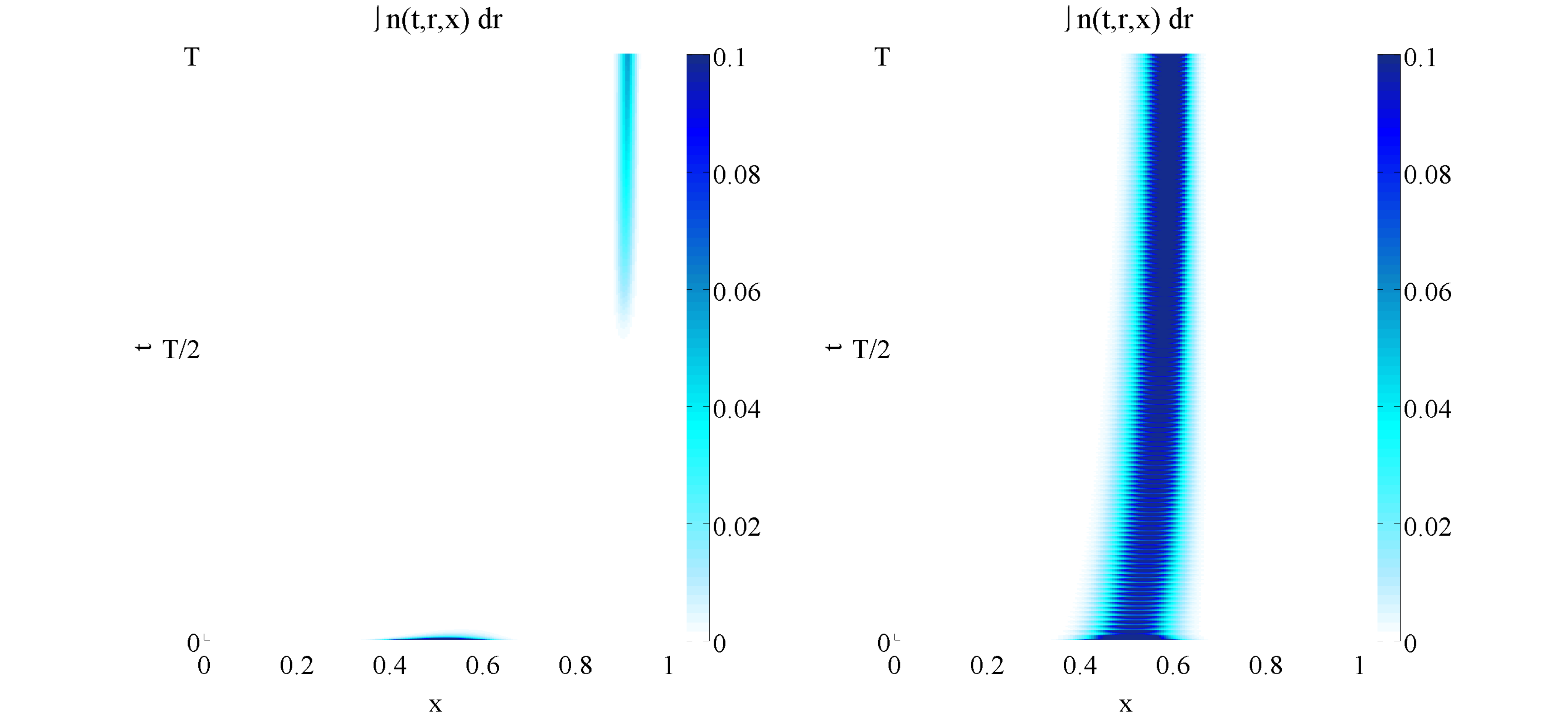}}\\
\subfigure[]{\includegraphics[width=0.8\textwidth]{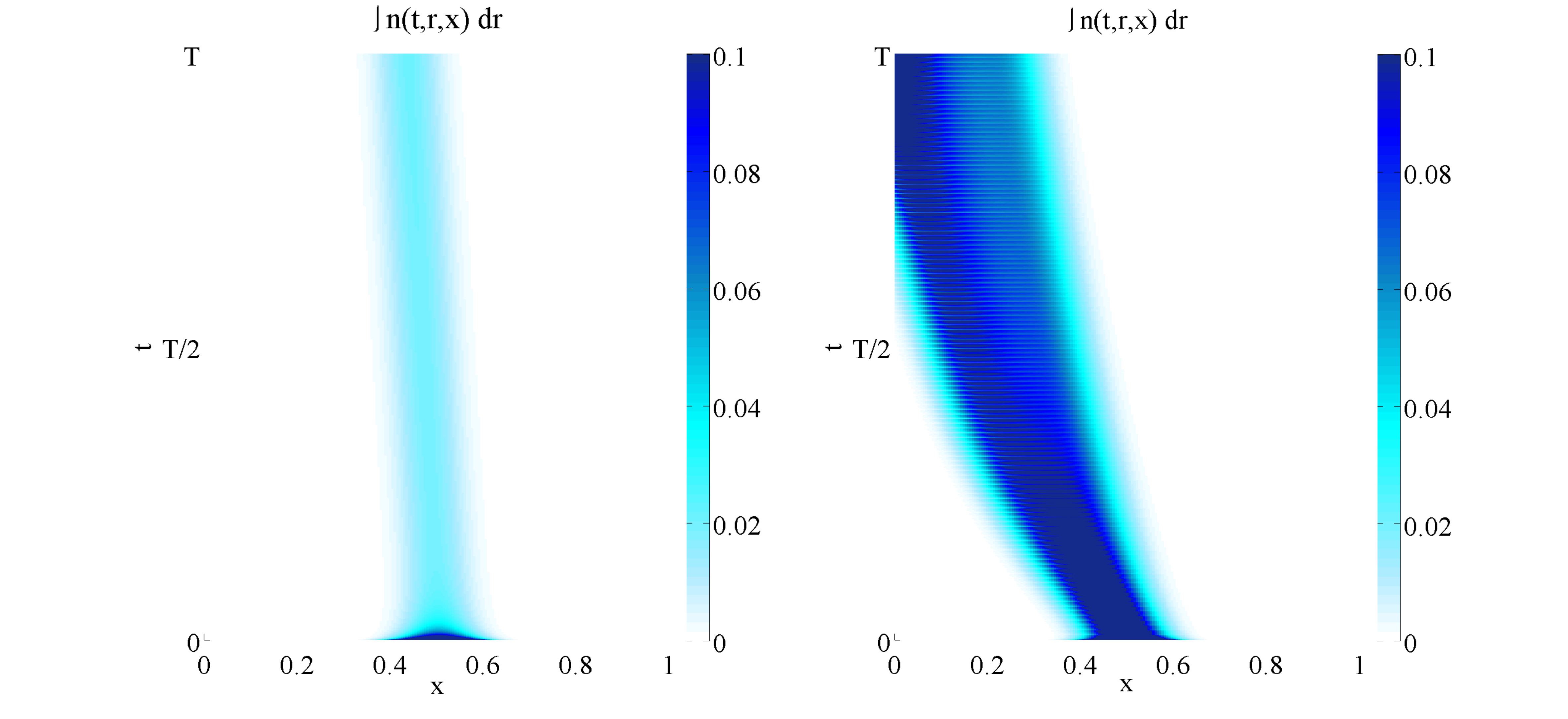}}
\caption{\textbf{(Constant vs bang-bang infusion of cytotoxic OR cytostatic drugs)}\\ 
\textbf{(a) Constant vs bang-bang infusion of cytotoxic drugs only.} Plots of $\int_0^1 n(t,r,x) dr$ for $C_2(t):=0$ and $C_1(t)$ defined as the $*$-line in the left panel in Fig.~\ref{IS} with $C_a=2$ (left panel) or the fixed line in the center panel in the same figure with $C_b=16$ (right panel).  Analogous results hold for bang-bang regimes illustrated in Fig.~\ref{IS} by the dashed line with $C_c=8$ and the dashed-dotted line with $C_d=4$ (data not shown). Constant infusion of cytotoxic drugs leads to a temporary reduction of the cancer cell density, while bang-bang delivery tends to slow down the evolution toward total resistance. \\ 
\textbf{(b) Constant vs bang-bang infusion of cytostatic drugs only.} Plots of $\int_0^1 n(t,r,x) dr$ for $C_1(t):=0$ and $C_2(t)$ defined as the $*$-line in the left panel in Fig.~\ref{IS} with $C_a=2$ (left panel) or the fixed line in the center panel in the same figure with $C_b=16$ (right panel).  Analogous results hold for bang-bang regimes illustrated in Fig.~\ref{IS} by the dashed line with $C_c=8$ and the dashed-dotted line with $C_d=4$ (data not shown). Constant infusion of cytostatic drugs slows down the selection of highly proliferative cancer clones, while bang-bang infusion weakly affects the dynamics of cancer cells with respect to the case without therapies.}\label{F23TN}
\end{figure}

Constant infusion of cytotoxic drugs leads to a temporary reduction of the cancer cell density, which is then followed by a relapse caused by the emergence of resistance (see left panel (a) in Fig.~\ref{F23TN}). On the other hand, the bang-bang infusion scheme with the same total dose slows down the selection of resistant cancer clones, but it is less effective in reducing the size of the tumor cell population (see right panel (a) in Fig.~\ref{F23TN}).

Then, we develop the same kind of analysis for cytostatic drugs only. We perform simulations with $C_1(t):=0$ and $C_2(t)$ defined as the $*$-line in the left panel in Fig.~\ref{IS} with $C_a=2$, or the fixed line in the center panel in Fig.~\ref{IS} with $C_b=16$, or the dashed line with $C_c=8$ or the dashed-dotted line with $C_d=4$. 

\begin{figure}[h!]
\centerline{\includegraphics[width=0.6\textwidth]{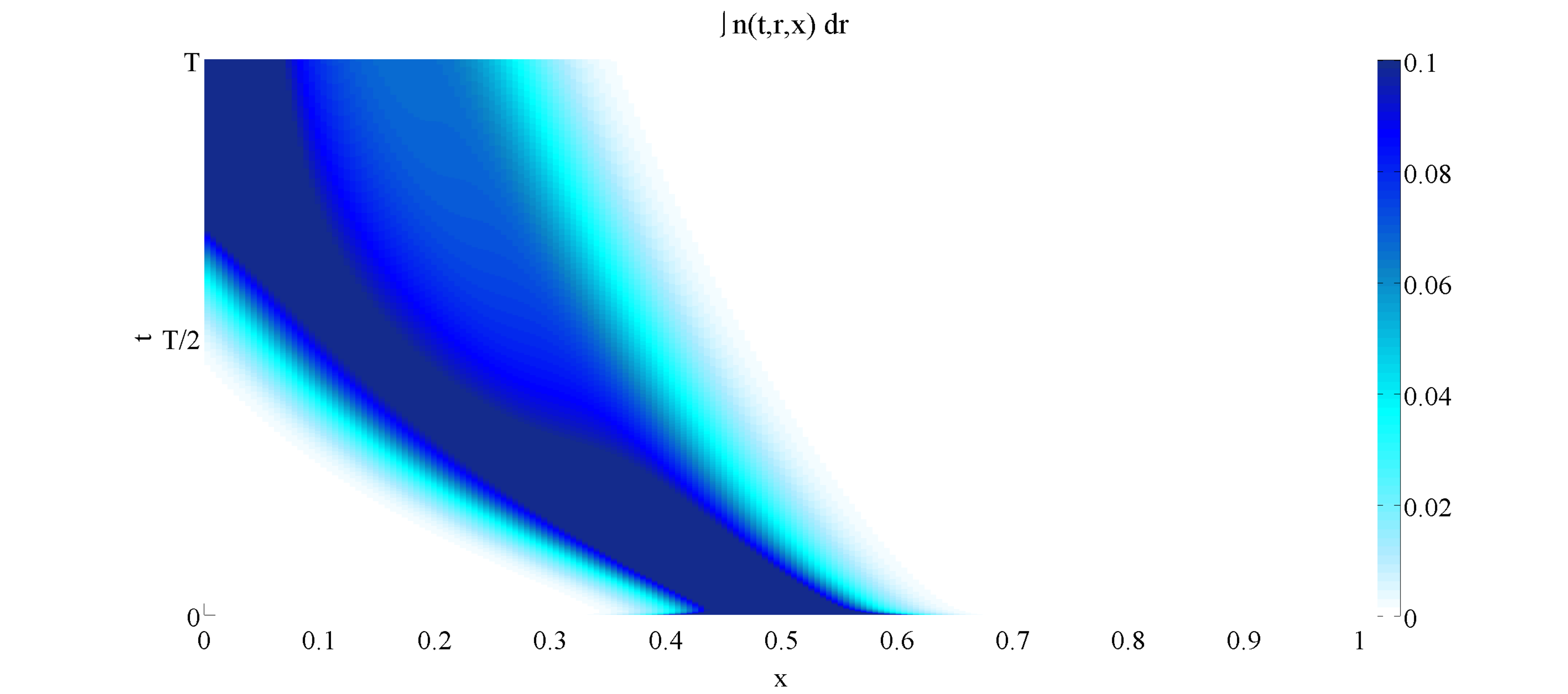}}
\caption{\label{F1TN}  \textbf{(Cell dynamics without therapies)} Plot of $\int_0^1 n(t,r,x) dr$ for $C_{1,2}(t):=0$. In agreement with the results presented in Section 3, cells characterized by a low expression level of resistance to cytotoxic therapies (i.e., a strong proliferative potential) are selected and intra-tumor heterogeneity is high. To be compared with Fig.~\ref{F23TN} and Fig.~\ref{F4TN}.}
\end{figure}

As we already know from Section 3, constant infusion of cytostatic drugs tends to slow down the evolution toward total sensitivity (i.e., the selection of high proliferative potentials) by comparison with the case without drugs (see left panel (b) in Fig.~\ref{F23TN} and compare it with Fig.~\ref{F1TN}). On the other hand, the dynamics of cancer cells under bang-bang delivery of cytostatic drugs is qualitatively the same as the one observed in the absence of therapies (see right panel (b) in Fig.~\ref{F23TN} and compare it with Fig.~\ref{F1TN}).
\\

{\it In conclusion to this section, we notice that constant infusion of cytotoxic drugs leads to a temporary reduction of the cancer cell density, while bang-bang delivery tends to slow down the evolution toward total resistance. On the other hand, bang-bang infusion of cytostatic drugs weakly affects the dynamics of cancer cells by comparison with the case without therapies. With the doses used in our tests, neither constant nor bang-bang infusion of cytotoxic/cytostatic drugs only allows a complete eradication of cancer cells.}

\subsection{Constant vs bang-bang infusion of cytotoxic AND cytostatic drugs}
\label{sec:41}
This subsection aims at making a comparison between the therapeutic effects of constant and bang-bang delivery of cytotoxic and cytostatic drugs in combination. Therefore, we perform simulations with $C_{1,2}(t)$ defined as the $*$-line in the left panel in Fig.~\ref{IS} with $C_a=1$ or the fixed line in the center panel in the same figure with $C_b=8$, or the dashed line with $C_c=4$ or the dashed-dotted line with $C_d=2$.

\begin{figure}[h!]
\centerline{\includegraphics[width=0.8\textwidth]{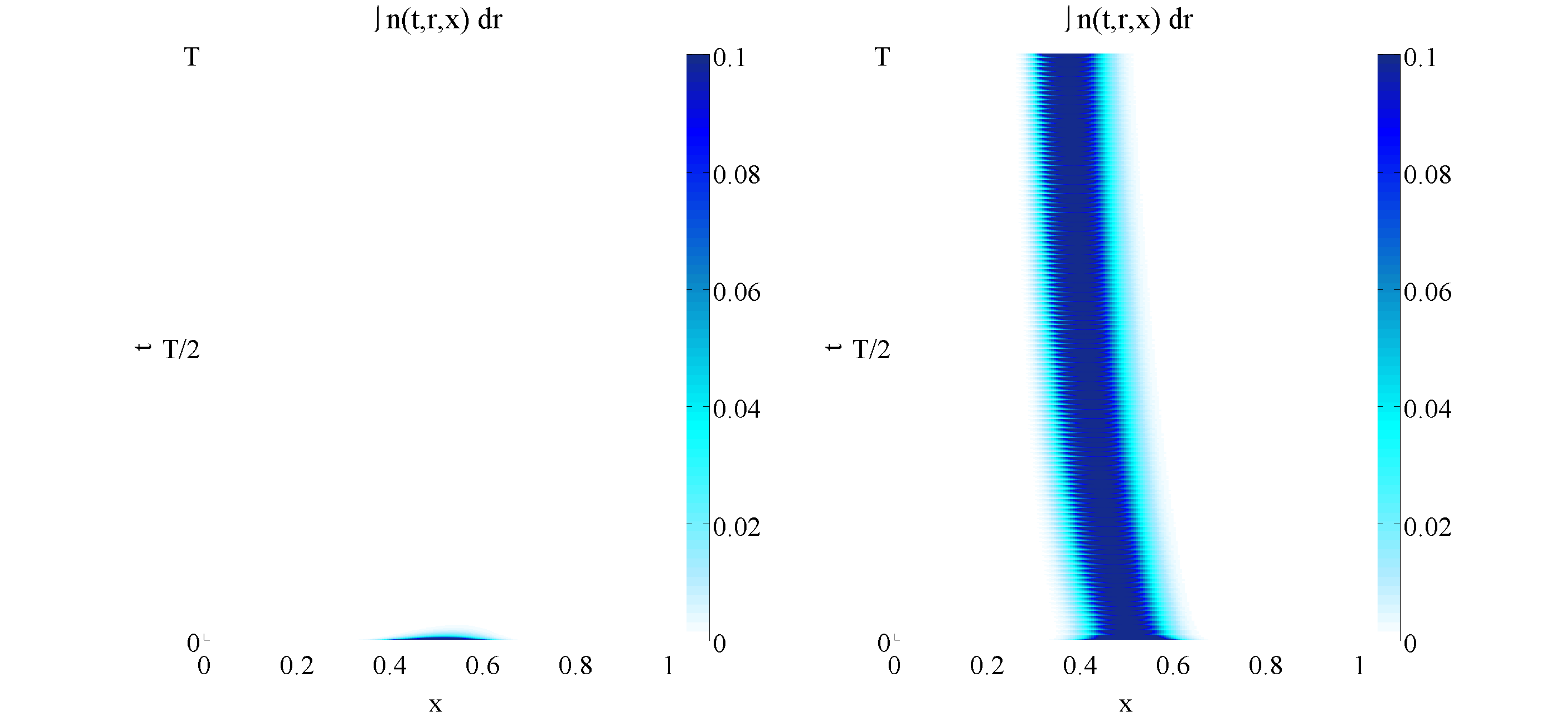}}
\caption{\label{F4TN}  \textbf{(Constant vs bang-bang infusion of cytotoxic AND cytostatic drugs)} Plots of $\int_0^1 n(t,r,x) dr$ for $C_{1,2}(t)$ defined as the $*$-line in the left panel in Fig.~\ref{IS} with $C_a=1$ (left panel) or the fixed line in the center panel of the same figure with $C_b=8$ (right panel). Analogous results hold for bang-bang regimes illustrated in Fig.~\ref{IS} by the dashed line with $C_c=4$ and the dashed-dotted line with $C_d=2$ (data not shown). While bang-bang infusion slows down the evolution toward total sensitivity which is observed in the absence of therapeutic agents, the constant infusion scheme pushes cancer cells toward extinction. }
\end{figure}

While bang-bang infusion slows down the evolution toward total sensitivity which is observed in the absence of therapeutic agents (see right panel in Fig.~\ref{F4TN} and compare it with Fig.~\ref{F1TN}), the constant infusion scheme at hand pushes cancer cells toward extinction (see left panel in Fig.~\ref{F4TN}). This is consistent with experimental observations suggesting that combination therapies can be more effective \cite{Janjigian2011,SilvaGatenby2010,Tabernero2007,Ye2013} and leads us to conclude that, keeping equal the total delivered dose of drugs, if cytotoxic and cytostatic drugs are used in combination, protocols relying on simultaneous bang-bang infusion can be less effective than protocols relying on simultaneous constant infusion.

It is worth noting that the total amount of delivered drugs is here the same as the ones considered in Subsection \ref{sec:41n}. Therefore, in agreement with the conclusions drawn in \cite{LorzLorenziHochbergClairambaultPerthame2012}, these results also suggest that looking for protocols based on different therapeutic agents in combination is a more effective strategy for fighting cancer rather than using high drug doses.
\\

{\it In conclusion to this section, we observe that effective anti-cancer treatments can be designed by making use of proper combinations between cytotoxic and cytostatic drugs. If these drugs are delivered together, constant supply is more effective than bang-bang infusion and can favor the total eradication of cancer cells.}

\subsection{Mixed constant/bang-bang infusions of cytotoxic AND cytostatic drugs}
\label{sec:43}
Finally, using the two types of drugs at hand in combination, we test the effects of delivery schedules relying on constant infusion of cytotoxic drugs and bang-bang infusion of cytostatic drugs, and vice-versa. Therefore, we perform simulations with $C_{1,2}(t)$ as in the right panel in Fig.~\ref{IS}. We set $C_{1}(t):=C_a=1$ and define $C_{2}(t)$ as the fixed line with $C_b=8$ (or the dashed line with $C_c=4$, or the dashed-dotted line with $C_d=2$), and vice-versa.

\begin{figure}[h!]
\centerline{\includegraphics[width=0.8\textwidth]{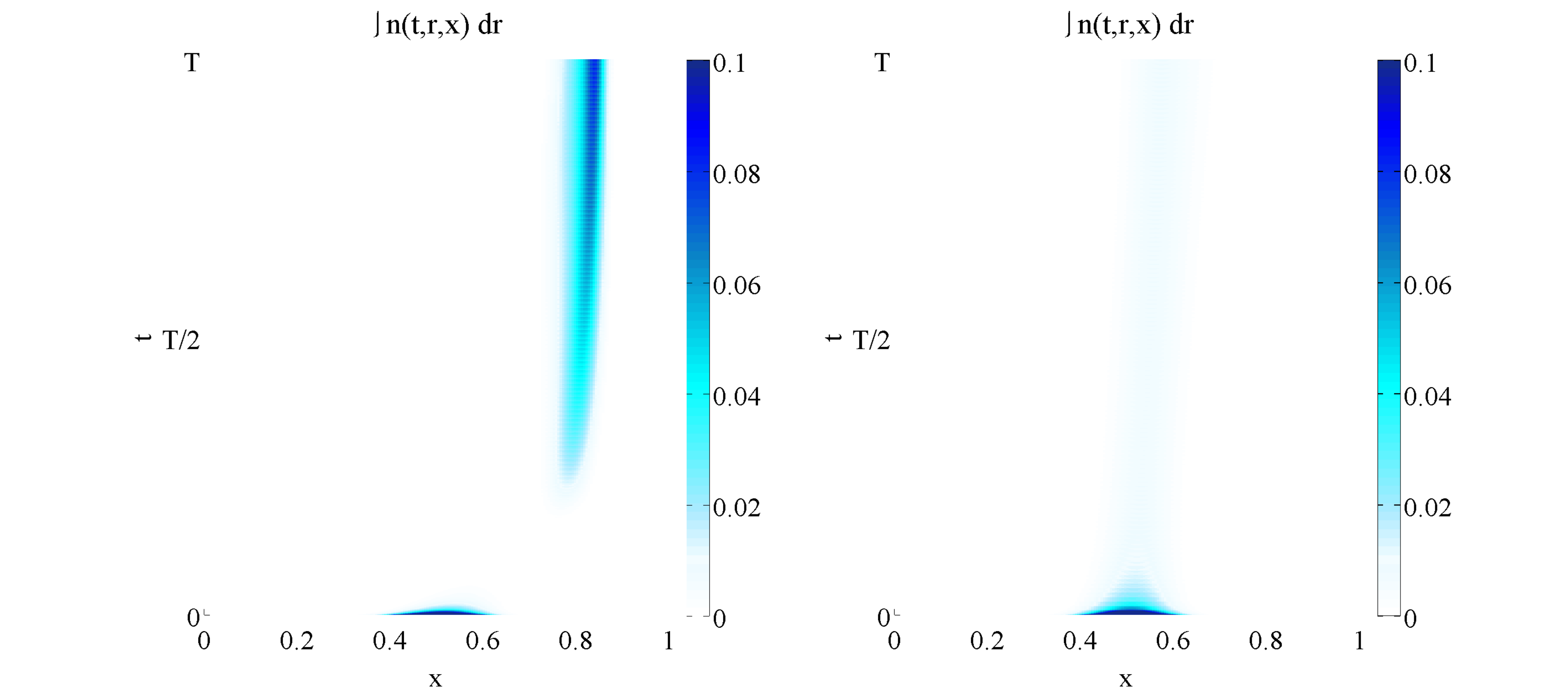}}
\caption{\label{F5TN}  \textbf{(Mixed constant/bang-bang infusions of cytotoxic AND cytostatic drugs)} Plots of $\int_0^1 n(t,r,x) dr$ for $C_{1,2}(t)$ as in the right panel in Fig.~\ref{IS} with $C_1(t):=C_a=1$ and $C_2(t)$ defined as the fixed line with $C_b=8$ (left panel), or vice-versa (right panel). Analogous results hold for bang-bang regimes illustrated in Fig.~\ref{IS} by the dashed line with $C_c=4$ and the dashed-dotted line with $C_d=2$ (data not shown). Bang-bang infusion of cytostatic drugs together with constant infusion of cytotoxic drugs causes a temporary reduction of the cancer cell density. On the other hand, bang-bang infusion of cytotoxic drugs together with constant delivery of cytostatic drugs can keep cancer cells close to extinction, although a detectable number of cancer cells survives within the population.}
\end{figure}

Bang-bang infusion of cytostatic drugs and constant infusion of cytotoxic drugs causes a temporary reduction of the cancer cell density (see left panel in Fig.~\ref{F5TN}). On the other hand, in good qualitative agreement with experimental observations \cite{FooChmielecki_etal2012}, therapeutic protocols relying on bang-bang infusion of cytotoxic drugs and constant delivery of cytostatic drugs can keep cancer cells close to extinction (see right panel in Fig.~\ref{F5TN}), although a detectable number of cancer cells survives within the population. 

These results, together with the ones presented in Subsection \ref{sec:41}, support the idea that more effective therapeutic protocols can be designed by using cytotoxic and cytostatic drugs in combination, with constant delivery for both drugs, or bang-bang infusion for cytotoxic drugs and constant infusion for cytostatic drugs. Moreover, if the delivered doses of each class of therapeutic agents are kept the same, protocols that make use of constant delivery for both classes of anti-cancer agents make possible a complete eradication of cancer cells, while protocols relying on bang-bang infusion of cytotoxic drugs and constant infusion of cytostatic drugs make only possible a good control on tumor size.
\\

{\it In conclusion to this section, we notice that therapeutic protocols relying on bang-bang infusion of cytotoxic drugs - constant delivery of cytostatic drugs are more effective than therapeutic protocols based on bang-bang infusion of cytostatic drugs - constant delivery of cytotoxic drugs. The former allow a good control on tumor size by keeping cancer cells close to extinction, while the latter make only possible a temporary reduction of the cancer cell density and leave space for tumor relapse, which arises due to the emergence of resistance.}

\clearpage

\section{Conclusions and perspectives}
Departing from theories derived in other contexts of population biology and Darwinian evolution, we have developed a structured population model for the dynamics of cancer cells exposed to cytotoxic and cytostatic drugs. Relying on the assumption that cells are organized in a radially symmetric spheroid, the present model takes explicitly into account the dynamics of resources and anti-cancer drugs, which define the cellular environment. In the present model, space structure together with  diffusion of nutrients and therapeutic agents are the key ingredients providing the basis for intra-tumor heterogeneity (i.e., the simultaneous selection of several levels of resistance/proliferative potential within the cancer cell population).
\subsection{Study of cell environmental adaptation and phenotypic heterogeneity}
In the framework of this model, we have first made use of numerical simulations to analyze the evolution of phenotypic heterogeneity and the emergence of resistance to therapies (see Section 3), and we have reached the following conclusions:
\\\\
$\bullet$ Phenotypic heterogeneity within solid tumor aggregates might be explained, at least partially, by cell adaptation to local conditions. In fact, cells characterized by different levels of proliferative potential and resistance to therapies are selected depending on space position, in relation with the distributions of resources  and anti-cancer drugs.\\
$\bullet$ Cytostatic drugs tend to slow down tumor evolution, while cytotoxic drugs favor the selection of highly resistant cancer clones and cause a decrease in the heterogeneity with respect to the resistance trait. In the framework of our model, this is not in contradiction with the Gause competitive exclusion principle. 

\subsection{Study of optimized therapeutic protocols}
As a second step, we have tested, \emph{in silico}, the capability of different therapeutic protocols to effectively contrast cancer progression. The cases of constant infusion, bang-bang delivery and mixed constant - bang-bang infusion of the anti-cancer drugs at hand have been considered (see Section 4). The results we have obtained lead us to the following conclusions:
\\\\
$\bullet$  Looking for combination therapies relying on cytotoxic and cytostatic drugs is a more effective strategy for fighting cancer rather than using high doses of cytotoxic or cytostatic drugs only.
\\
$\bullet$  If cytotoxic and cytostatic drugs are delivered together, constant supply is more effective than bang-bang infusion, or mixed constant and bang-bang infusion, since it can actually push cancer cells toward extinction. 
\\
$\bullet$ Therapeutic protocols relying on bang-bang infusion of cytotoxic drugs and constant delivery of cytostatic drugs favor a good control on tumor size and they are more effective than therapeutic protocols based on bang-bang infusion of cytostatic drugs and constant delivery of cytotoxic drugs. Since constant supplies might be excluded in practice for toxicity reasons, bang-bang cytotoxic associated with constant cytostatic infusion might turn out to be optimal.

\subsection{Perspectives}
Future researches will be addressed to extend the present model in order to include the dynamics of healthy cells and studying how to handle unwanted toxic side effects related to the delivery of anti-cancer agents in healthy cell populations. In view of this, a possible research direction is to approach the toxicity problem at stake as an optimal control problem, where the systemic cost for human body related to cancer growth and therapy infusion has to be minimized under the constraint that a minimal density of healthy cells should be preserved inside the system. In the same way, in view of adaptive therapy \cite{GatenbySilvaGillies_2009}, preserving a minimal proportion of cancer cells under a sensitivity threshold $x\leq x_s$ (to be tuned) is another possible constraint to be represented in an optimal control problem. This general direction of research, using optimal control settings, clearly aims at therapeutic optimization in the clinic of cancers.

From a modeling point of view, the present theoretical study, that has taken into account (additionally to the space variable $r$) only one drug resistance trait ($x$) for two different classes of anticancer drugs should also be completed in the future by further work involving a multidimensional structure variable $x$ including different resistant phenotypes to different drugs, and possibly other phenotypes related to epithelial-to-mesenchymal transition, glycolytic metabolism phenotype, dormancy, etc. as cell subpopulations less accessible to treatment than the classical forms of \emph{in situ} proliferating epithelial solid tumors.

From a more biological point of view, understanding what are the cell population characters (genetically or epigenetically determined?) aggregated in this structure variable $(r,x)$ standing for spatial heterogeneity and quantitated drug resistance, and how it can be related in experimental measurements with gene expression or epigenetic enzyme activity, is a big challenge that must be faced in transdisciplinary studies gathering mathematicians and biologists. We know from Luria and Delbr\"uck's {\it princeps} experiment \cite{LuriaDelbruck1943} that some drug resistance, due to stochastic genetic mutations in cell populations, occurring prior to drug exposure, is likely to exist in cancer cell populations, all the more so as genome instability is a common feature of these cells. Finding out what are the respective parts played in drug resistance by purely stochastic processes \cite{Gupta2011} on the one hand, and by more Lamarckian phenomena resulting from adaptation of the cells surviving a massive drug insult, involving epigenetic mechanisms in response \cite{Sharma2010}, and thus justifying the use of deterministic models, is another challenge that we intend to tackle in forthcoming studies.

\bigskip

\noindent{\em Acknowledgments.} T.L. was supported by the Fondation Sciences Math\'ematiques de Paris and the FIRB project - RBID08PP3J

\bibliography{AABJT}

\begin{thebibliography}{10}

\bibitem{BuschXingYu_etal2009}
{\sc T.~M. Busch, X.~Xing, G.~Yu, A.~Yodh, E.~P. Wileyto, H.-W. Wang,
  T.~Durduran, T.~C. Zhu, and K.~K.-H. Wang}, {\em Fluence rate-dependent
  intratumor heterogeneity in physiologic and cytotoxic responses to photofrin
  photodynamic therapy.}, Photochem Photobiol Sci, 8 (2009), pp.~1683--1693.

\bibitem{FooChmielecki_etal2012}
{\sc J.~Foo, J.~Chmielecki, W.~Pao, and F.~Michor}, {\em Effects of
  pharmacokinetic processes and varied dosing schedules on the dynamics of
  acquired resistance to erlotinib in egfr-mutant lung cancer}, J Thorac Oncol,
  7 (2012), pp.~1583--1593.

\bibitem{Gatenby_2009}
{\sc R.~Gatenby}, {\em A change of strategy in the war on cancer}, Nature, 459
  (2009), pp.~508--509.

\bibitem{GatenbySilvaGillies_2009}
{\sc R.~Gatenby, A.~Silva, R.~Gillies, and B.~Frieden}, {\em Adaptive therapy},
  Cancer Research, 69 (2009), pp.~4894--4903.

\bibitem{GerlingerRowanHorswell_etal2012}
{\sc M.~Gerlinger, A.~J. Rowan, S.~Horswell, J.~Larkin, D.~Endesfelder,
  E.~Gronroos, P.~Martinez, and et~al.}, {\em Intratumor heterogeneity and
  branched evolution revealed by multiregion sequencing}, N Engl J Med, 366
  (2012), pp.~883--892.

\bibitem{Gerlinger2010}
{\sc M.~Gerlinger and C.~Swanton}, {\em How darwinian models inform therapeutic
  failure initiated by clonal heterogeneity in cancer medicine.}, Br J Cancer,
  103 (2010), pp.~1139--1143.

\bibitem{GoldieColdman1998}
{\sc J.~Goldie and A.~Coldman}, {\em Drug resistance in cancer: mechanisms and
  models}, Cambridge University Press, 1998.

\bibitem{Gottesman2002}
{\sc M.~Gottesman}, {\em Mechanisms of cancer drug resistance}, Annu Rev Med,
  53 (2002), pp.~615--627.

\bibitem{Gupta2011}
{\sc P.~B. Gupta, C.~M. Fillmore, G.~Jiang, S.~D. Shapira, K.~Tao,
  C.~Kuperwasser, and E.~S. Lander}, {\em Stochastic state transitions give
  rise to phenotypic equilibrium in populations of cancer cells.}, Cell, 146
  (2011), pp.~633--644.

\bibitem{Janjigian2011}
{\sc Y.~Y. Janjigian, C.~G. Azzoli, L.~M. Krug, L.~K. Pereira, N.~A. Rizvi,
  M.~C. Pietanza, M.~G. Kris, M.~S. Ginsberg, W.~Pao, V.~A. Miller, and G.~J.
  Riely}, {\em Phase i/ii trial of cetuximab and erlotinib in patients with
  lung adenocarcinoma and acquired resistance to erlotinib.}, Clin Cancer Res,
  17 (2011), pp.~2521--2527.

\bibitem{KimmelSwierniak2006}
{\sc M.~Kimmel and A.~\'Swierniak}, {\em Tutorials in Mathematical Biosciences
  III}, Springer, 2006, ch.~Control theory approach to cancer chemotherapy:
  benefiting from phase dependence and overcoming drug resistance,
  pp.~185--221.

\bibitem{KomarovaWodarz2005}
{\sc N.~Komarova and D.~Wodarz}, {\em Drug resistance in cancer: Principles of
  emergence and prevention}, Proc Natl Acad Sci USA, 102 (2005),
  pp.~9714--9719.

\bibitem{lavi2013role}
{\sc O.~Lavi, J.~M. Greene, D.~Levy, and M.~M. Gottesman}, {\em The role of
  cell density and intratumoral heterogeneity in multidrug resistance}, Cancer
  research, 73 (2013), pp.~7168--7175.

\bibitem{LedzewiczSchattler2002}
{\sc U.~\L\k{e}d\.zewicz and H.~Sch\"attler}, {\em Optimal bang-bang controls
  for a two-compartment model in cancer chemotherapy}, Jour Opt Theory and
  Appl, 114 (2002), pp.~609--637.

\bibitem{LorzLorenziHochbergClairambaultPerthame2012}
{\sc A.~Lorz, T.~Lorenzi, M.~E. Hochberg, J.~Clairambault, and B.~Perthame},
  {\em Populational adaptive evolution, chemotherapeutic resistance and
  multiple anti-cancer therapies}, ESAIM: Mathematical Modelling and Numerical
  Analysis, 47 (2013), pp.~377--399.

\bibitem{LorzMirrahimiPerthame2010}
{\sc A.~Lorz, S.~Mirrahimi, and B.~Perthame}, {\em Dirac mass dynamics in a
  multidimensional nonlocal parabolic equation}, Comm Partial Differential
  Equations, 36 (2011), pp.~1071--1098.

\bibitem{LuriaDelbruck1943}
{\sc S.~E. Luria and M.~Delbr\"uck}, {\em Mutations of bacteria from virus
  sensitivity to virus resistance.}, Genetics, 28 (1943), pp.~491--511.

\bibitem{MerloPepperReid2006}
{\sc L.~Merlo, J.~Pepper, B.~Reid, and C.~Maley}, {\em Cancer as an
  evolutionary and ecological process}, Nat Rev Cancer, 6 (2006), pp.~924--935.

\bibitem{MirrahimiRaoul2013}
{\sc S.~Mirrahimi and G.~Raoul}, {\em Dynamics of sexual populations structured
  by a space variable and a phenotypical trait}, Theor Popul Biol, 84 (2013),
  pp.~87--103.

\bibitem{Mitchison2012}
{\sc T.~Mitchison}, {\em The proliferation rate paradox in antimitotic
  chemotherapy}, Mol Biol Cell, 23 (2012), pp.~1--6.

\bibitem{GB.BP:08}
{\sc B.~Perthame and G.~Barles}, {\em Dirac concentrations in
  {L}otka-{V}olterra parabolic {PDE}s}, Indiana Univ. Math. J., 57 (2008),
  pp.~3275--3301.

\bibitem{Scotto2003}
{\sc K.~Scotto}, {\em Transcriptional regulation of abc drug transporters},
  Oncogene, 22 (2003), pp.~7496--7511.

\bibitem{Sharma2010}
{\sc S.~V. Sharma, D.~Y. Lee, B.~Li, M.~P. Quinlan, F.~Takahashi,
  S.~Maheswaran, U.~McDermott, N.~Azizian, L.~Zou, M.~A. Fischbach, K.-K. Wong,
  K.~Brandstetter, B.~Wittner, S.~Ramaswamy, M.~Classon, and J.~Settleman},
  {\em A chromatin-mediated reversible drug-tolerant state in cancer cell
  subpopulations.}, Cell, 141 (2010), pp.~69--80.

\bibitem{SilvaGatenby2010}
{\sc A.~Silva and R.~Gatenby}, {\em A theoretical quantitative model for
  evolution of cancer chemotherapy resistance}, Biology Direct, 22 (2010),
  pp.~5--25.

\bibitem{Swanton2012}
{\sc C.~Swanton}, {\em Intratumor heterogeneity: evolution through space and
  time}, Cancer Res, 72 (2010), pp.~4875--4882.

\bibitem{Szakacs2006}
{\sc G.~Szak\'acs, G., J.~K. Paterson, J.~A. Ludwig, C.~Booth-Genthe, and M.~M.
  Gottesman}, {\em Targeting multidrug resistance in cancer.}, Nat Rev Drug
  Discov, 5 (2006), pp.~219--234.

\bibitem{Tabernero2007}
{\sc J.~Tabernero, E.~{Van Cutsem}, E.~D\'iaz-Rubio, A.~Cervantes, Y.~Humblet,
  T.~Andr\'e, J.-L. {Van Laethem}, P.~Souli\'e, E.~Casado, C.~Verslype, J.~S.
  Valera, G.~Tortora, F.~Ciardiello, O.~Kisker, and A.~{de Gramont}}, {\em
  Phase ii trial of cetuximab in combination with fluorouracil, leucovorin, and
  oxaliplatin in the first-line treatment of metastatic colorectal cancer.}, J
  Clin Oncol, 25 (2007), pp.~5225--5232.

\bibitem{TomasettiLevy2010}
{\sc C.~Tomasetti and D.~Levy}, {\em An elementary approach to modeling drug
  resistance in cancer}, Math Biosci Eng, 7 (2010), pp.~905--918.

\bibitem{TomasettiLevy22010}
\leavevmode\vrule height 2pt depth -1.6pt width 23pt, {\em SBEC 2010, IFMBE
  Proceedings 32}, Springer, Berlin, 2010, ch.~Drug resistance always depends
  on the cancer turnover rate, pp.~552--555.

\bibitem{Weinberg2007}
{\sc R.~Weinberg}, {\em The Biology of Cancer}, Garland Science, 2007.

\bibitem{Ye2013}
{\sc L.-C. Ye, T.-S. Liu, L.~Ren, Y.~Wei, D.-X. Zhu, S.-Y. Zai, Q.-H. Ye,
  Y.~Yu, B.~Xu, X.-Y. Qin, and J.~Xu}, {\em Randomized controlled trial of
  cetuximab plus chemotherapy for patients with kras wild-type unresectable
  colorectal liver-limited metastases.}, J Clin Oncol, 31 (2013),
  pp.~1931--1938.

\bibitem{Zhou1996}
{\sc D.~C. Zhou, S.~Ramond, F.~Vigui\'e, A.~M. Faussat, R.~Zittoun, and J.~P.
  Marie}, {\em Sequential emergence of mrp- and mdr1-gene over-expression as
  well as mdr1-gene translocation in homoharringtonine-selected k562 human
  leukemia cell lines.}, Int J Cancer, 65 (1996), pp.~365--371.

\end{thebibliography}
\bibliographystyle{siam}

\end{document}